\documentclass[manuscript]{aastex62}
\usepackage{textcomp}
\usepackage{url}
\usepackage{placeins}
\usepackage{multirow}
\usepackage{graphicx}
\usepackage{amsmath}
\usepackage{amssymb,latexsym}
\usepackage{threeparttable}
\usepackage{longtable}
\usepackage[caption=false]{subfig}
\usepackage{ulem} 
\usepackage{natbib}

\hypersetup{
     colorlinks   = true,
     citecolor     = blue
} 

\newcommand{\bjdtdb}{\ensuremath{\rm {BJD_{TDB}}}}

\newcommand{\msun}{\ensuremath{\,M_\Sun}}
\newcommand{\rsun}{\ensuremath{\,R_\Sun}}
\newcommand{\lsun}{\ensuremath{\,L_\Sun}}

\newcommand{\mj}{\ensuremath{\,M_{\rm J}}}
\newcommand{\rj}{\ensuremath{\,R_{\rm J}}}

\newcommand{\fave}{\langle F \rangle}
\newcommand{\fluxcgs}{10$^9$ erg s$^{-1}$ cm$^{-2}$}
\newcommand{\mos}{\,m\,s$^{-1}$}

\newcommand\msini{\ifmmode{{\mathrm M} \sin i}\else${{\mathrm M} \sin i}$\fi}

\def\emphasize#1{{\sl#1\/}}
\def\arg#1{{\it#1\/}}

\def\edcomment#1{\iffalse\marginpar{\raggedright\sl#1\/}\else\relax\fi}
\reversemarginpar
\tabletypesize{\scriptsize}
\usepackage{apjfonts}

\begin{document}
\title{{\sc Minerva}-Australis I: Design, Commissioning, \& First Photometric Results}
\author{Brett Addison}
\affil{University of Southern Queensland, Centre for Astrophysics, West Street, Toowoomba, QLD 4350 Australia} 
\author{Duncan J. Wright}
\affil{University of Southern Queensland, Centre for Astrophysics, West Street, Toowoomba, QLD 4350 Australia}
\author{Robert A. Wittenmyer}
\affil{University of Southern Queensland, Centre for Astrophysics, West Street, Toowoomba, QLD 4350 Australia}
\author{Jonathan Horner}
\affil{University of Southern Queensland, Centre for Astrophysics, West Street, Toowoomba, QLD 4350 Australia}
\author{Matthew W. Mengel}
\affil{University of Southern Queensland, Centre for Astrophysics, West Street, Toowoomba, QLD 4350 Australia}
\author{Daniel Johns}
\affil{Department of Physical Sciences, Kutztown University, Kutztown, PA 19530, USA}
\author{Connor Marti}
\affil{Department of Astronomy, Williams College, 33 Lab Campus Drive, Williamstown, MA 01267 USA}
\author{Belinda Nicholson}
\affil{University of Southern Queensland, Centre for Astrophysics, West Street, Toowoomba, QLD 4350 Australia}
\author{Jack Okumura}
\affil{University of Southern Queensland, Centre for Astrophysics, West Street, Toowoomba, QLD 4350 Australia}
\author{Brendan Bowler}
\affil{Department of Astronomy, The University of Texas at Austin, TX 78712, USA}
\author{Ian Crossfield}
\affil{Department of Physics, Massachusetts Institute of Technology, Cambridge, MA, USA}
\author{Stephen R. Kane}
\affil{Department of Earth Sciences, University of California, Riverside, CA 92521, USA}
\author{John Kielkopf}
\affil{Department of Physics and Astronomy, University of Louisville, Louisville, KY 40292, USA}
\author{Peter Plavchan}
\affil{Department of Physics \& Astronomy, George Mason University, 4400 University Drive MS 3F3, Fairfax, VA 22030, USA}
\author{C.G. Tinney}
\affil{Exoplanetary Science at UNSW, School of Physics, UNSW Sydney, NSW 2052, Australia}
\author{Hui Zhang}
\affil{School of Astronomy and Space Science, Key Laboratory of Modern Astronomy and Astrophysics in Ministry of Education,
Nanjing University, Nanjing 210046, Jiangsu, China}
\author{Jake T. Clark}
\affil{University of Southern Queensland, Centre for Astrophysics, West Street, Toowoomba, QLD 4350 Australia}
\author{Mathieu Clerte}
\affil{University of Southern Queensland, Centre for Astrophysics, West Street, Toowoomba, QLD 4350 Australia}
\author{Jason D. Eastman}
\affil{Center for Astrophysics, Harvard \& Smithsonian, 60 Garden St., Cambridge, MA 02138, USA}
\author{Jon Swift}
\affil{The Thacher School, 5025 Thacher Road, Ojai, CA 93023, USA}
\author{Michael Bottom}
\affil{California Institute of Technology, 1200 E California Blvd, Pasadena, CA 91125, USA}
\author{Philip Muirhead}
\affil{Department of Astronomy, Institute for Astrophysical Research, Boston University, 725 Commonwealth Avenue, Boston, MA 02215,
USA}
\author{Nate McCrady}
\affil{University of Montana, Department of Physics and Astronomy, 32 Campus Drive, No. 1080, Missoula, Montana 59812, USA}
\author{Erich Herzig}
\affil{California Institute of Technology, 1200 E California Blvd, Pasadena, CA 91125, USA}
\author{Kristina Hogstrom}
\affil{California Institute of Technology, 1200 E California Blvd, Pasadena, CA 91125, USA}
\author{Maurice Wilson}
\affil{Center for Astrophysics, Harvard \& Smithsonian, 60 Garden St., Cambridge, MA 02138, USA}
\author{David Sliski}
\affil{Department of Physics and Astronomy, University of Pennsylvania, Philadelphia, PA 19104, USA}
\author{Samson A. Johnson}
\affil{Department of Astronomy, The Ohio State University, 140 West 18th Avenue, Columbus, OH 43210, USA}
\author{Jason T. Wright}
\affil{Department of Astronomy \& Astrophysics, 525 Davey Laboratory
The Pennsylvania State University, University Park, PA, 16802, USA}
\affil{Center for Exoplanets and Habitable Worlds, 525 Davey Laboratory
The Pennsylvania State University, University Park, PA, 16802, USA}
\author{Cullen Blake}
\affil{Department of Physics and Astronomy, University of Pennsylvania, Philadelphia, PA 19104, USA}
\author{Reed Riddle}
\affil{California Institute of Technology, 1200 E California Blvd, Pasadena, CA 91125, USA}
\author{Brian Lin}
\affil{California Institute of Technology, 1200 E California Blvd, Pasadena, CA 91125, USA}
\author{Matthew Cornachione}
\affil{Department of Physics, United States Naval Academy, 572C Holloway Rd, Annapolis, MD 21402, USA}
\author{Timothy R. Bedding}
\affil{Sydney Institute for Astronomy (SIfA), School of Physics, University of Sydney, NSW 2006, Australia}
\affil{Stellar Astrophysics Centre, Department of Physics and Astronomy, Aarhus University, Ny Munkegade 120, DK-8000 Aarhus C, Denmark}
\author{Dennis Stello}
\affil{School of Physics, University of New South Wales, NSW 2052, Australia}
\author{Daniel Huber}
\affil{Institute for Astronomy, University of Hawaii, 2680 Woodlawn Drive, Honolulu, HI 96822, USA}
\author{Stephen Marsden}
\affil{University of Southern Queensland, Centre for Astrophysics, West Street, Toowoomba, QLD 4350 Australia}
\author{Bradley D. Carter}
\affil{University of Southern Queensland, Centre for Astrophysics, West Street, Toowoomba, QLD 4350 Australia}

\submitjournal{Publications of the Astronomical Society of the Pacific}
\published{November 2019}

\begin{abstract}
The {\sc Minerva}-Australis telescope array is a facility dedicated to the follow-up, confirmation, characterization, and mass measurement of planets orbiting bright stars discovered by the \textit{Transiting Exoplanet Survey Satellite} ({\textit {TESS}}) -- a category in which it is almost unique in the Southern Hemisphere. It is located at the University of Southern Queensland's Mount Kent Observatory near Toowoomba, Australia. Its flexible design enables multiple 0.7\,m robotic telescopes to be used both in combination, and independently, for high-resolution spectroscopy and precision photometry of {\textit{TESS}} transit planet candidates. {\sc Minerva}-Australis also enables complementary studies of exoplanet spin-orbit alignments via Doppler observations of the Rossiter-McLaughlin effect, radial velocity searches for non-transiting planets, planet searches using transit timing variations, and ephemeris refinement for \textit{TESS} planets. In this first paper, we describe the design, photometric instrumentation, software, and science goals of {\sc Minerva}-Australis, and note key differences from its Northern Hemisphere counterpart, the {\sc Minerva} array. We use recent transit observations of four planets, WASP-2b, WASP-44b, WASP-45b, and HD\,189733b, to demonstrate the photometric capabilities of {\sc Minerva}-Australis.

\end{abstract}

\section{Introduction}
There has long been interest in the discovery of planets around other stars. Early attempts to find such worlds, however, got off to a slow and rocky start with several exoplanetary detection claims being either later retracted, or never confirmed, such as the proposed planets orbiting 70 Ophiuchi \citep{70Oph}, Barnard's Star, \citep{1963AJ.....68..515V,1995Ap&SS.223...91G}, and the pulsar PSR B1829-10 \citep{Bailes,Lyne}). It took until the announcement of the first confirmed exoplanet orbiting a Sun-like star in 1995 \citep[51 Pegasi b; ][]{1995Natur.378..355M} to truly kick off the ``Exoplanet Era.''

In the years immediately following that discovery, the number of confirmed exoplanets grew slowly. As we have become ever more adept at finding new planets, however, the number known has grown exponentially, especially over the last decade. This is due, in large part, to the extremely successful {\textit {Kepler}} mission launched by NASA in 2009 \citep[][]{2010ApJ...713L..79K} to search for planets via their transits. The spacecraft's four year primary mission, together with its more recent K2 program extension \citep[][]{2014PASP..126..398H}, confirmed the existence of over 2500 planets\footnote{See \url{https://exoplanetarchive.ipac.caltech.edu/} for the latest tally.}, including many that resemble nothing found in the solar system.

This incredible diversity includes the so-called 'hot Jupiters' and 'hot Neptunes' \citep[e.g.,][]{1995Natur.378..355M,2000ApJ...529L..45C,2007A&A...472L..13G,2010ApJ...710.1724B,2013AJ....146..113B}, planets moving on extremely eccentric orbits \citep[e.g.,][]{2006MNRAS.369..249J,2017AJ....154..274W},  planets with densities greater than iron and even osmium \citep[e.g.,][]{2008A&A...491..889D,2014ApJ...789..154D,2018arXiv180804533J}, or comparable to styrofoam \citep[e.g.,][]{2011A&A...531A..40F,2015ApJ...809...26W,2017AJ....153..215P}. Perhaps most surprisingly, {\textit{Kepler}} revealed that planets between the size of Earth and Neptune (``super-Earths'' or ``mini-Neptunes'') are incredibly common, despite the fact that no analog exists in the solar system \citep[e.g.,][]{2009Natur.462..891C,2018A&A...612A..95B}.

The primary goals of {\textit{Kepler}} were to perform a detailed exoplanet census and to measure the frequency distribution function for planets around other stars. This was accomplished by continually monitoring $\sim$150,000 stars in the northern constellation of Cygnus for transits \citep[][]{2010ApJ...713L..79K} for a period in excess of four years. Chief among {\textit {Kepler}}'s results is the revelation that planets are ubiquitous, and that the majority of stars host small planets, with mini-Neptunes and super-Earths being the most common of those found on orbits of $\leq200$ days \citep[][]{2013ApJ...766...81F}. {\textit{Kepler}} also revealed that Earth-sized planets ($0.5\leq R_{P} \leq1.4 R_{\oplus}$) are particularly common around cool stars ($\mathrm{T_{eff}}\leq4000$\,K), with an occurrence rate of just over 50\% \citep{2013ApJ...767...95D}. Indeed, based on {\textit{Kepler}} data, Dressing \& Charbonneau estimated the occurrence rate of Earth-size planets in the habitable zone as $0.15^{+0.13}_{-0.06}$ planets per cool star. This suggest that the nearest transiting Earth-size planet in the habitable zone could be located within 21\,pc of Earth.

Despite the stunning success of the {\textit {Kepler}} mission, little is known about the compositions, masses, and densities of the majority of the {\textit {Kepler}} planets. The reason for this is that the majority of the planet-hosting stars identified by {\textit{Kepler}} are either too faint for further follow-up investigations using existing facilities, or would require an inordinate investment of time on large telescopes. Because of the significant resources that are required to convert the large number of {\textit {Kepler}} candidates into confirmed planets and measure their masses, only about $50\%$ of {\textit {Kepler}}'s candidate planets have been confirmed, and of these, only $\sim10\%$ have mass measurements\footnote{Determined using the NASA Exoplanet Archive (\url{https://exoplanetarchive.ipac.caltech.edu/}). There are 2347 confirmed {\textit{Kepler}} planets and 244 of them have mass measurements listed in the table.}. 

On 2018 April 18, NASA launched its next-generation exoplanet finder, the \textit{Transiting Exoplanet Survey Satellite} \citep[{\textit {TESS}};][]{2015JATIS...1a4003R}. Unlike {\textit {Kepler}}, which observed a single small region on the sky, {\textit {TESS}} expands the search for planets to nearly the entire sky. {\textit {TESS}} consists of four wide-angle cameras that each have a field of view of $24^{\circ}\times24^{\circ}$, yielding a total field of view for {\textit {TESS}} of $96^{\circ}\times24^{\circ}$. The spacecraft is oriented such that one of the cameras is centered on one of the ecliptic poles while the others are pointed progressively closer to the ecliptic. {\textit {TESS}} will monitor each $24^{\circ}$ wide strip on the celestial sphere for a period of 27 days before moving on to an adjacent strip of the sky. As such, the majority of stars will be observed for 27 days, while those closer to the ecliptic poles will be observed for longer. As a result of this strategy, stars within $\sim 12^\circ$ of the ecliptic poles will be observed for a year. {\textit {TESS}} will observe the southern ecliptic hemisphere in its first year of operation before moving on to the northern ecliptic hemisphere in the second year of its mission. 

Throughout the course of its initial two-year mission, {\textit {TESS}} will survey approximately 200,000 of the brightest stars in the sky with a cadence of two minutes. Planets discovered around these bright stars will be suitable for ground-based follow-up observations to both confirm their existence and facilitate their characterization \citep[e.g.][]{tesspaper, nontess}. Data will also be returned on an additional 20 million stars from ''full-frame images'', taken with a cadence of 30 minutes. As a result, there will be no shortage of planet candidates coming from {\textit {TESS}} that will need follow-up observations. Additionally, stars observed by {\textit {TESS}} will be, on average, a hundred times brighter than those observed by {\textit {Kepler}}, and it is expected that {\textit {TESS}} will deliver a yield of thousands of new planets orbiting bright stars.

With the expected flood of planet candidates being found by {\textit {TESS}} to be orbiting bright stars, dedicated facilities are urgently needed to confirm the candidates and characterize them. The radial velocity technique is the primary method to deliver the critical planetary parameters, such as mass and orbital eccentricity, that are required to properly characterize the planetary system. Most of the existing facilities capable of carrying out the required high-precision radial velocity measurements, however, are subject to intense competition and scheduling constraints (particularly on shared large telescopes). Traditionally, radial velocity programs are allocated blocks of time (a couple of weeks to a month) on large telescopes during bright nights (though some such as the Hobby-Eberly Telescope and WIYN are working to facilitate queue and cadence observations).

With the expected large number of planet candidates to be delivered by {\textit {TESS}}, the most exciting of which will be low-mass planets with orbital periods exceeding one month (in particular those planets orbiting within the habitable zone around M-dwarf stars), this strategy simply will not work. This is the primary cause of the significant bottlenecks experienced during the follow-up work carried out on {\textit {Kepler}} candidates \citep{2015AJ....149..143F}. 

Similarly with {\textit {TESS}}, we will be in a situation where we have too many planets, and too few telescopes to confirm them.

To address this bottleneck issue with {\textit {TESS}} follow-up, we are commissioning the {\sc Minerva}-Australis facility at the University of Southern Queensland's Mount Kent Observatory (MKO). {\sc Minerva}-Australis builds on the template and groundwork of a similar facility in the Northern Hemisphere called {\sc Minerva} (MINiature Radial Velocity Array) located at Mt. Hopkins in the Arizona desert \citep[][]{2015JATIS...1b7002S}. Whereas the primary goal of the northern {\sc Minerva} observatory is to search for small-mass planets orbiting nearby bright stars through high-cadence radial velocity observations, {\sc Minerva}-Australis will be primarily focused on supporting the follow-up work of NASA's {\textit {TESS}} mission.

The {\sc Minerva}-Australis collaboration consists of the University of Southern Queensland as the primary investigator along with the University of Texas at Austin, Massachusetts Institute of Technology, University of California at Riverside, University of Louisville, George Mason University, the University of New South Wales, University of Florida, and Nanjing University as co-investigators and major funding partners in this project. It is also a participating member of the \textit{TESS} Follow-up Observing Program (TFOP\footnote{\url{https://tess.mit.edu/followup/}}) Working Group (WG). The primary goal of TFOP WG is to coordinate the follow-up observations of \textit{TESS} planet candidates to measure masses for 50 transiting planets smaller than four Earth radii. Additionally, TFOP WG is fostering communication and coordination within its network of participants and community at large to optimize the follow-up work of \textit{TESS} planet candidates and minimize wasteful duplication of observations and analysis. {\sc Minerva}-Australis is primarily involved in two of the TFOP Sub Groups (SGs), SG2 for reconnaissance spectroscopy and SG4 for precision radial velocity work. 

Secondary science objectives for the {\sc Minerva}-Australis project include the measurement of the spin-orbit alignment of planetary systems through radial velocity and Doppler tomography observations of the Rossiter--McLaughlin effect \citep{1924ApJ....60...15R,1924ApJ....60...22M,2000A&A...359L..13Q,2014ApJ...790...30J,2018arXiv180900314A}. Spin--orbit alignments can provide key insights into the formation and migration histories of exoplanets \citep[e.g., see,][]{1996Natur.380..606L,2000Icar..143....2B,2008ApJ...686..621F,2011Natur.473..187N,2011ApJ...735..109W}, in particular hot \citep{CridaBatygin:2014,2015ARA&A..53..409W} and warm Jupiters \citep{2014ApJ...781L...5D}, and compact transiting multi-planet systems \citep{2013ApJ...771...11A,2018AJ....155...70W}. Additionally, Doppler tomography observations can aid in the confirmation of transiting planet candidates orbiting rapidly rotating stars that are not amenable to precise radial velocity observations \citep{2014ApJ...790...30J}. Another science goal is to carry out long-term radial velocity monitoring of planetary systems found by {\textit {TESS}}. Such observations could reveal the existence of non-transiting long-period planets that can provide constraints on the migration history of the inner transiting planet(s) \citep[e.g.][]{nt1,nt2,nt3}. 

While spectroscopy and radial velocity observations are the primary focus of {\sc Minerva}-Australis, high-precision and high-cadence photometry is also an important component of this project. 

Fluctuations observed in the out-of-transit photometry of a star can be used to disentangle the radial velocity variations that are produced by stellar activity, such as from starspots and from the suppression of convective blueshift occurring in active regions on the stellar surface, from planetary signals \citep{2011A&A...528A...4B}. This is particularly the case when photometry and radial velocity data are obtained simultaneously \citep{2014MNRAS.443.2517H}, as can be done with {\sc Minerva}-Australis. Such data will provide a better understanding of the effects of stellar activity on radial velocity observations, enabling the detection of sub-Neptune mass planets and more accurate determinations of their masses.

Simultaneous photometry is also useful when carrying out Rossiter--McLaughlin effect observations. These observations are typically carried out several hundred orbital periods after the last published transit in the literature and the ephemerides have usually become out of date. Simultaneous photometry can be used to lock down the transit ingress, mid, and egress times needed for properly analyzing Rossiter--McLaughlin data. Additionally, stellar activity can deform the Rossiter--McLaughlin signal, which can cause significant variations in the measured spin-orbit angle (up to $\sim42^{\circ}$) from transit to transit \citep{2018A&A...619A.150O}. Simultaneous photometry can provide information about the properties of the active regions on a stellar surface that will allow for better modeling of the Rossiter--McLaughlin signal \citep{2018A&A...619A.150O}.

While it is expected the majority of transiting planets detected by {\textit {TESS}} will come from the pre-selected bright stars with a two minute cadence, undoubtedly some transit detections will come from the full-frame images that are taken at a 30 minute cadence. These planets will require follow-up transit photometry to help improve the transit parameters such as the orbital period. {\sc Minerva}-Australis will be utilized for this task as such planet candidates are found. We will also be using photometry to search for transit timing variations through photometric transit observations that could reveal the presence of additional planets \citep{tt1,tt2} as well as transit observations to keep the transit ephemerides up-to-date for {\textit {TESS}} planets.

Other ancillary science goals include observations of predicted solar system occultation events. Such observations have yielded improved information on the size, shape, and albedo of small solar system objects \citep{2011Natur.478..493S,2011AJ....141...67S,2012Natur.491..566O} as well as led to the discovery of ring systems around some of these minor bodies \citep{2014Natur.508...72B,2017Natur.550..219O}. We are also planning on continuing the radial velocity follow-up of targets that were originally observed as part of the Anglo-Austrian Planet Search program \citep[AAPS; e.g.,][]{2011ApJ...732...31T,2014ApJ...783..103W} with the aim of extending the radial velocity data-set baseline to enable the detection of longer orbital period ($\geq5$\,yr) planets. By continuing the AAPS legacy survey, our goal is to expand the population of known Jupiter or Saturn analogs and determine the degree to which the solar system is unusual or unique.

We have organized the paper as follows: Section~\ref{MKO} provides an overview of the MKO and the reasons for selecting the site for {\sc Minerva}-Australis. Section~\ref{facilities} describes the {\sc Minerva}-Australis facilities and hardware, including the telescopes and enclosures, spectrograph, camera, and control building. In Section~\ref{FirstResults}, we present the first science results from follow-up photometric transit observations of four transiting planets, including WASP-2b, WASP-44b, WASP-45b, and HD 189733b. In Section~\ref{Conclusions} we summarize the {\sc Minerva}-Australis facility, goals, and future work.

\section{Mount Kent Observatory}
\label{MKO}
\begin{figure}
	\centering
	\includegraphics[width=1.0\linewidth]{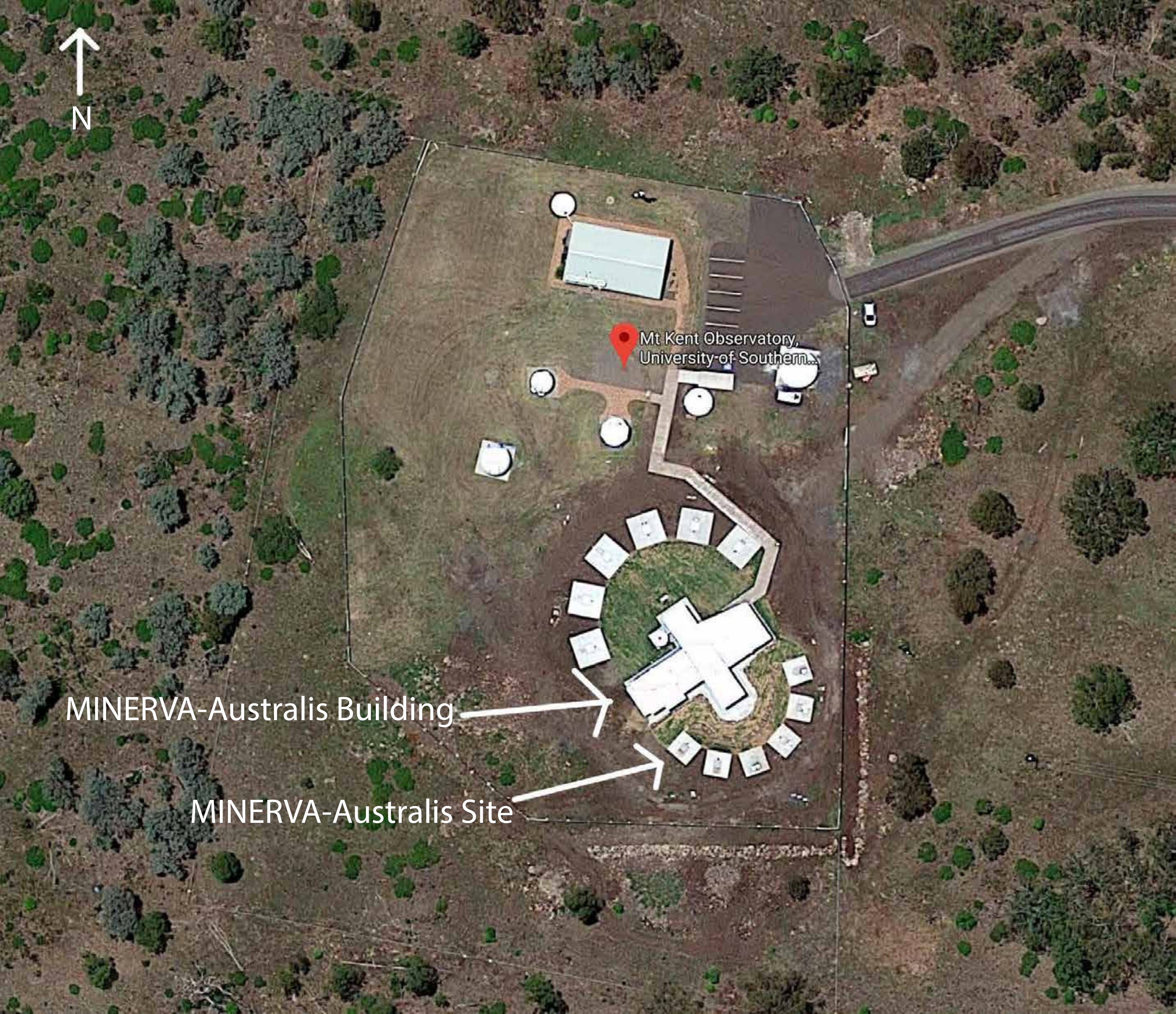}
	\caption{Google Maps image of the Mount Kent Observatory showing the location of the {\sc Minerva}-Australis telescopes and building.}
	\label{fig:Google_Map_Minerva}
\end{figure}

The {\sc Minerva}-Australis facility is being commissioned at the MKO, located in the Darling Downs in Queensland, Australia, approximately 25\,km south-southwest of Toowoomba and 120\,km west-southwest of Brisbane. It is situated at an altitude of $\sim680$\,m and a latitude and longitude of  27\degr47\arcmin53\arcsec\,S and 151\degr51\arcmin20\arcsec\,E. The site already houses three telescopes of the Shared Skies project, operated jointly between the University of Louisville, KY, and the University of Southern Queensland: the 0.1\,m aperture wide field OMara robotic telescope, used for education, and two PlaneWave Instrument telescopes (a CDK20 and a CDK700; CDK: corrected Dall-Kirkham) used for KELT{\footnote{The Kilodegree Extremely Little Telescope survey \citep[e.g.][]{Kelt1,Kelt2,JackKELT}}} and {\textit {TESS}} precision exoplanet transit photometry follow-up. Figure~\ref{fig:Google_Map_Minerva} shows a Google Maps image of the {\sc Minerva}-Australis telescope sites, building location, and other facilities on the MKO.

The Mount Kent site was selected based on good weather conditions with an average of $\sim$296 clear/mostly clear nights per year, reasonably good seeing conditions (estimated to be around 1.6\arcsec\ from seeing measurements reported by the facilities at the site), and existing facilities and support. The weather data have been obtained from the Bureau of Meteorology website\footnote{\url{http://www.bom.gov.au/}} using the Oakey Aero weather station (a nearby weather station with 35 years of historic climate data) located about 44\,km from Mount Kent. Conditions at the Oakey weather station should serve as a good proxy for the conditions observed at Mount Kent and give a good estimate of the number of usable nights. Therefore, we anticipate $\sim296$\footnote{This time excludes additional factors like unusually high humidity, dust storms, and maintenance.} nights per year with observable weather conditions with at least an average of 6.7 usable hours and a median seeing of 1.6\arcsec. 

\section{Facilities}
\label{facilities}

In this section, we give details of the {\sc Minerva}-Australis hardware, highlight the ways in which our new facility differs from the northern {\sc Minerva} facility \citep[][]{2015JATIS...1b7002S}, and discuss the reasoning behind those choices. 

\subsection{Telescopes and Enclosures}

\begin{figure}
	\centering
	\includegraphics[width=0.80\linewidth]{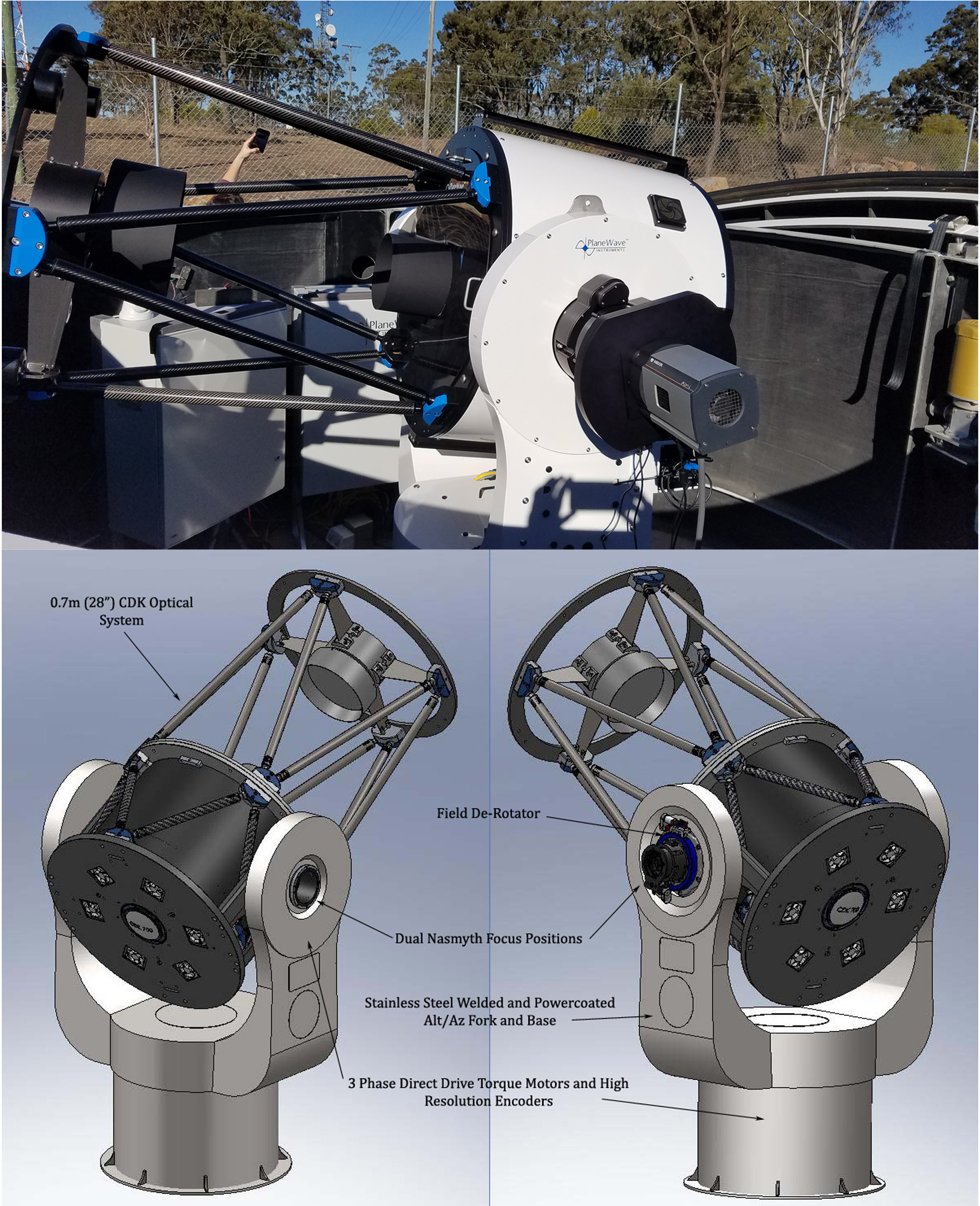}
	\caption{Top: {\sc Minerva}-Australis PlaneWave CDK-700 telescope inside an AstroHaven Enterprises Dome. Bottom: schematic of the PlaneWave CDK-700 telescope obtained from the PlaneWave website.}
	\label{fig:Telescope}
\end{figure}

{\sc Minerva}-Australis will comprise up to six independently operated 0.7\,m PlaneWave CDK-700 altitude/azimuth mounted telescopes{\footnote{{\url{http://planewave.com/products-page/cdk700}}, three installed as of 2019 February}} (see Figure~\ref{fig:Telescope}), arranged in a semi-circle, all feeding light to a single Kiwispec\footnote{\url{https://www.kiwistaroptics.com/}} high-resolution spectrograph \citep{2012SPIE.8446E..88B}. In contrast, northern {\sc Minerva} comprises four 0.7\,m PlaneWave CDK-700 telescopes (equivalent to a single 1.4\,m diameter aperture versus a single 1.7\,m aperture for {\sc Minerva}-Australis). The additional two telescopes in the {\sc Minerva}-Australis array provide us with $1.5\times$ the collecting area and $\sim1.2\times$ increase in signal-to-noise ratio over northern {\sc Minerva}. The PlaneWave CDK-700 telescope has a compact design that is 2.4\,m tall when pointed at zenith and a radius of maximum extent of 1.5\,m. The telescopes use a CDK optical setup to remove off-axis coma, astigmatism, and field curvature. Over a three minute interval, the telescopes have a pointing accuracy of 10\arcsec\, RMS, pointing precision of 2\arcsec, and a tracking accuracy of 1\arcsec. The telescopes are controlled through a PlaneWave Interface. They also have a very fast slew rate of 15\degr\, per second, enabling the telescopes to slew between any two points on the sky within 10 seconds. For a complete discussion of the PlaneWave CDK-700 telescopes hardware and specifications, we refer the reader to the northern {\sc Minerva} facility publication \citep{2015JATIS...1b7002S}. Here we provide a summary of the important aspects of the telescopes and list the specifications in Table~\ref{table:CDK_700_specs}.

\begin{figure}[ht]
	\centering
	\includegraphics[width=1.0\linewidth]{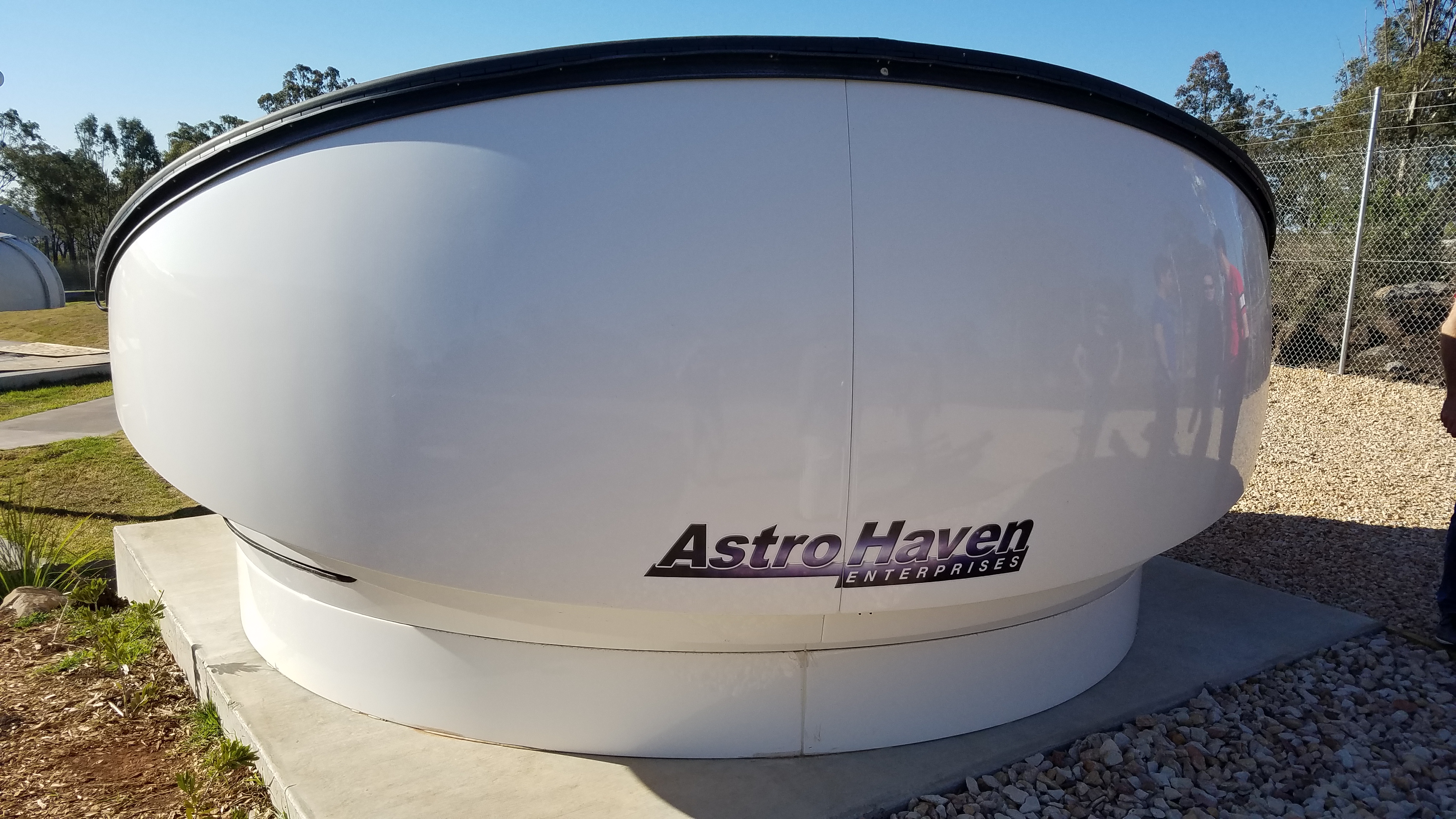}
	\caption{AstroHaven Enterprises dome housing a single PlaneWave CDK-700 Telescope.}
	\label{fig:dome}
\end{figure}

\begin{center}
\begin{deluxetable}{p{5cm}p{9cm}}[ht]
\tabletypesize{\scriptsize}
\tablecaption{CDK-700 Specifications}
\startdata \\ [-6.0ex]
\multicolumn{2}{c}{\bf{Optical System}} \\
\hline \\ [-8.0ex]
Optical design\dotfill        & CDK \\ 
Aperture\dotfill              & 700\,mm (27.56 \,in) \\
Focal length\dotfill          & 4540\,mm \\
Focal ratio\dotfill           & 6.5  \\
Central obscuration\dotfill   & 47\% primary diameter \\
Back focus\dotfill            & 305\,mm from mounting surface \\
Focus position\dotfill        & Nasmyth (dual) \\
Dimensions\dotfill            & $93.73^{\prime\prime}$\,H $\times$ $43.25^{\prime\prime}$\,W $\times$ $39^{\prime\prime}$\,D \\
Weight\dotfill                & 1200\,lbs \\
Optical performance\dotfill   & 1.8\,$\mu$m RMS spot size on axis \\
Image scale\dotfill           & 22\,$\mu$m per arcsecond \\
Optimal field of view\dotfill & 70\,mm (0.86 degrees) \\
Fully baffled field\dotfill   & 60\,mm \\ [+2.0ex]
\hline \\ [-8.0ex]
\multicolumn{2}{c}{\bf{Mechanical Structure}} \\
\hline \\ [-8.0ex]
Mount\dotfill                 & Altitude-azimuth \\
Fork\dotfill                  & Monolithic U-shaped fork arm \\
Azimuth bearing\dotfill       & 20\,in diameter thrust bearing \\
Altitude bearing\dotfill       & $2 \times 8.5$\,in OD ball bearings \\
Optical tube\dotfill           & Dual truss structure \\ [+2.0ex]
\hline \\ [-8.0ex]
\multicolumn{2}{c}{\bf{Motion Control}} \\
\hline \\ [-8.0ex]
Motors\dotfill                 & Direct drive, three-phase axial flux torque motor \\
Encoders\dotfill               & Stainless steel encoder tape with 81\,mas resolution \\
Motor torque\dotfill           & $\sim 35$\,ft-lbs \\
Slew rate\dotfill              & $15^\circ$\,s$^{-1}$ \\ [+2.0ex]
\hline \\ [-8.0ex]
\multicolumn{2}{c}{\bf{System Performance}} \\
\hline \\ [-8.0ex]
Pointing accuracy\dotfill      & $10^{\prime\prime}$ RMS \\
Pointing precision\dotfill     & $2^{\prime\prime}$ RMS \\
Tracking accuracy\dotfill      & $1^{\prime\prime}$ RMS over three minutes \\
Field de-rotator\dotfill       & $3\,\mu$m peak-to-peak 35\,mm off axis over one hour \\
\enddata 
\end{deluxetable}
\label{table:CDK_700_specs}
\end{center}

Each of the PlaneWave CDK-700 telescopes is housed in its own AstroHaven Enterprises 12.5\,ft (3.81\,m) diameter dome\footnote{\url{https://www.astrohaven.com/}}, as shown in Figure~\ref{fig:dome}. The dome is designed for remote and robotic operations from anywhere in the world. It can fully open, giving full access to the sky, and can achieve thermal equilibrium very rapidly, reducing the effects of ``dome seeing''. The dome can also open rapidly, in approximately 20\,s, with each hemisphere opening independently. In contrast, the northern {\sc Minerva} uses an "Aqawan" enclosure that was developed by Las Cumbres Observatory engineer Annie Kirby.  A rectangular Aqawan can house two PlaneWave CDK-700 telescopes whereas each AstroHaven dome for {\sc Minerva}-Australis houses one Planewave telescope that can be independently operated.

The AstroHaven domes are powered by 240\,V/15\,A three-phase power which is converted to 24V DC within its control panel and the telescopes by a 240\,V/10\,A power supply. This is then connected to an internal uninterruptible power supply (UPS) and stand-by generator, ensuring the dome can close if the site loses power and we can cease observing until power is restored. Communication to the domes are established by a TCP/IP interface and are controlled through ASCII string commands. We are currently in the process of implementing an automated dome closure protocol in case communication is lost to the domes and telescopes for remote observing carried out in the future. 

A web camera is situated inside each of the domes to provide a live video feed to the user. A weather station is located approximately 10\,m to the north of the {\sc Minerva}-Australis building which provides real-time temperature, humidity, wind speed, and wind direction measurements. In addition, it has a rain sensor that will alert the user to rain and send a signal to control to close up the dome.

\subsection{Spectrograph}
All of the telescopes in the {\sc Minerva}-Australis array simultaneously feed a single Kiwispec R4-100 high-resolution spectrograph \citep{2012SPIE.8446E..88B} via fiber optic cables. The specifications for the spectrograph are provided in Table~\ref{table:KiwiSpec_specs}. The spectrograph is bench-mounted and housed in an insulated, environmentally controlled enclosure. Kiwispec uses an R4 echelle for the primary dispersion while a volume phase holographic grism is used for the cross-dispersion. The fibers are aligned in the cross-dispersion direction of the spectrometer, and form seven individual echelle traces that are imaged on a 2k$ \times $2k detector. The detector has a wavelength coverage from 500 to 630\,nm over from 26 echelle orders with a resolution of $R\approx80,000$.

We currently use $50\mu$m circular fiber cables with a $70\mu$m cladding diameter that are butt-coupled to $50\mu$m circular fibers and a numerical aperture of 0.22 to feed scrambled light to the spectrograph. The five science fibers are bracketed by two additional calibration fibers that provide a simultaneous and stable thorium-argon wavelength calibration source. Octagonal fibers will replace the circular fibers in the final configuration of the instrument by mid-2019. Northern {\sc Minerva} in contrast uses octagonal fibers and has four science fibers for the four telescopes in the array. It also use an iodine absorption cell placed in the light path instead of a thorium-argon lamp for wavelength calibration source. While the Minerva design was optimized for iodine work, we currently use thorium-argon as it provides a good wavelength reference required for precision radial velocity work ($<1$\,\mos, \citealt{ 2003Msngr.114...20M}). Additional reasons for choosing thorium-argon over iodine cell include better throughput of the system and no contamination of the spectra from iodine absorption lines. More details on the commissioning and performance of the {\sc Minerva}-Australis Kiwispec spectrograph will be presented in a follow-up paper. 

\begin{center}
\begin{deluxetable}{p{5cm}p{9cm}}[ht]
\tabletypesize{\scriptsize}
\tablecaption{KiwiSpec R4-100 Specifications}
\startdata \\ [-6.0ex]
\multicolumn{2}{c}{\bf{Characteristics}} \\
\hline \\ [-8.0ex]
Spectral resolution\dotfill   & 80,000 \\ 
Wavelength range\dotfill      & 500\,nm -- 630\,nm \\
Echelle orders\dotfill        & 26 \\
Detector size\dotfill         & 2k$ \times $2k \\
Cross-disperser\dotfill       & Anamorphic VPH grisms \\
Beam diameter\dotfill         & 100mm (at echelle grating), 33mm (at cross-disperser) \\
Main fibers\dotfill           & $8\times$ $50\mu$m circular fibers (six science and two calibration) \\
Average sampling\dotfill      & 3.5 pixel per FWHM \\
Calibration\dotfill           & Simultaneous ThAr lamp \\
Environment for main optics\dotfill    & Vacuum operation, 1\,mK temperature stability \\
Environment for camera optics\dotfill  & Pressure sealed operation, 20\,mK temperature stability \\
Long-term instrument stability\dotfill  & Goal of $1$\,\mos \\
\enddata 
\end{deluxetable}
\label{table:KiwiSpec_specs}
\end{center}

\subsection{Photometric Camera} \label{sec:Photometric_Camera}
The {\sc Minerva}-Australis telescopes can each be equipped with an Andor iKon-L 936 camera\footnote{\url{https://andor.oxinst.com/products/ikon-xl-and-ikon-large-ccd-series/ikon-l-936}}, the specifications are provided in Table~\ref{table:camera_specs}. Switching between our standard spectroscopic mode to the photometric observing mode is done through a simple flipping of the telescope's M3 mirror in the PWI software interface to direct stellar light to the imagining camera instead of the fiber that feeds the KiwiSpec spectrograph.

As of 2018 September, one photometric camera is present on Telescope 1. Further cameras will be acquired subject to funding. The camera consists of $2048 \times 2048$ square $13.5\,\mu$m pixels that provides an on-sky field of view of 20.9\arcmin, contain a deep depletion sensor with fringe suppression (BEX2-DD), and have extended range dual anti-reflection (AR) coating. The deep depletion sensor enables the camera to be sensitive to light from the near-ultraviolet to the near-infrared ($1\,\mu$m) for precision photometry. The cameras are also equipped with a five-stage thermo-electric cooling system that allows the sensor to be cooled down to $-100\degr$\,C, keeping dark current to a minimum, without having to use liquid nitrogen.

Each telescope uses a CenterLine Color 10 Position Color Filter Wheel (FLI CL1-10 CFW), with second-generation Sloan \emphasize{g'2}, \emphasize{r'2}, \emphasize{i'2}, \emphasize{z'2} filters and a narrow-band H$\alpha$ filter (planned for a later date). We chose these standard sets of filters to provide us with flexibility for our transit observations while observing in variable conditions and for our other auxiliary programs. 

The Sloan \emphasize{r'2} filter is used for most transit photometric observations. For the transit observations of WASP-2b, WASP-44b, and WASP-45b, we used the \emphasize{r'2} filter. This filter provides good throughput while minimizing atmospheric extinction effects. We used the \emphasize{z'2} band filter for observing the transit of HD 189733b since the star is quite bright ($V=7.6$) and the quantum efficiency of our camera is a factor of 2 below peak. The \emphasize{z'2} filter is suitable for observing bright targets without reaching the non-linearity point or pixel saturation level for the detector at reasonable exposure lengths.

To obtain high-precision photometry at high cadence with reasonably short readout times, our observations are carried out using the $1.0$\,MHz pixel readout mode. The total readout time for the detector in this mode is $\sim4$\,s with a readout noise of $22.2$\,e$^{-}$.

\begin{center}
\begin{deluxetable}{p{5cm}p{9cm}}[ht]
\tabletypesize{\scriptsize}
\tablecaption{Andor iKon-L DEX2-DD Specifications}
\startdata \\ [-6.0ex]
\multicolumn{2}{c}{\bf{Characteristics}} \\
\hline \\ [-8.0ex]
Detector size\dotfill         & 2k$ \times $2k \\
Pixel size\dotfill            & $13.5\times13.5$\,$\mu$m \\
Image area\dotfill            & $27.6 \times 27.6$\,mm \\
On-sky field of view\dotfill            & $20.9$\,\arcmin \\
Pixel well depth\dotfill      & 150,000\,e$^{-}$ \\
Operating temperature\dotfill & $-100$\,$^{\circ}$C \\
Dark current (e$^{-}$\,pixel$^{-1}$\,s$^{-1}$)\dotfill     & 0.0003 \\
Pixel readout rates\dotfill   & 5.0, 3.0, 1.0, and 0.05 MHz \\
Read noise (e$^{-}$)\dotfill      & \  \\
\ \ \ \ 0.05\,MHz\dotfill          & 8.7 \\
\ \ \ \ 1\,MHz\dotfill             & 22.2 \\
\ \ \ \ 3\,MHz\dotfill             & 40.2 \\
\ \ \ \ 5\,MHz\dotfill             & 70.3 \\
Wavelength regions (quantum efficiency $\geq$ 50\%) \dotfill & 350--975\,nm \\
Peak quantum efficiency \dotfill & 750\,nm \\
\enddata 
\end{deluxetable}
\label{table:camera_specs}
\end{center}

\subsection{Control Building}

While {\sc Minerva}-Australis is primarily designed for automated observing, it can be controlled both on site and remotely. The \$2 million control facility features a purpose-built class 100,000 clean room that houses the spectrograph, with the critical components inside a vacuum chamber and thermally stabilized to $\pm0.01$\,K. Additionally, it contains an UPS room where power for the entire {\sc Minerva}-Australis facility is routed through for an uninterruptible power supply and a control room that houses the computers and network equipment.  

\section{First Science Results with {\sc Minerva}-Australis}
\label{FirstResults}

While the primary focus of {\sc Minerva}-Australis will be the radial velocity follow-up of transiting planet candidates found by {\textit {TESS}}, five secondary photometric science goals include: photometric follow-up of {\textit {TESS}} and other transit survey planets to ensure that the ephemerides are up-to-date and accurate for future follow-up observations \citep[e.g.,][]{2016AJ....151..137H,2018AJ....156..181W}, ruling out potential false positives from nearby eclipsing binaries \citep{2018arXiv180301869C,2018arXiv180610142Z},
searching for transit timing variations \citep{tt1,tt2} and longer-period planets \citep[e.g.,][]{nt1,nt3}, and follow-up of planets found by radial velocity observations from the AAPS \citep[e.g.,][]{2011ApJ...732...31T,2014ApJ...783..103W}. In addition, target-of-opportunity observations of high-priority solar system occultation events \citep[e.g.,][]{2011Natur.478..493S,2012Natur.491..566O,2017Natur.550..219O} are also planned using a very high-cadence camera that will be installed at a future date. With this in mind, we conducted high-precision photometry observations of four known transiting exoplanets as part of the commissioning operations and to benchmark our photometric precision.

\subsection{High-Precision Photometry}
\label{PrecisionPhotometry}

We carried out high-precision photometry for four known transiting planets exoplanets, WASP-2b, WASP-44b, WASP-45b, and HD 189733b using the first telescope that was installed at the {\sc Minerva}-Australis site. Photometry was obtained using the Andor iKON-L camera and the Sloan \emphasize{r'2} (WASP-2b, WASP-44b, and WASP-45b) and \emphasize{z'2} (HD 189733b) filters as discussed in section~\ref{sec:Photometric_Camera}. Maxim DL\footnote{\url{http://diffractionlimited.com/product/maxim-dl/}} was used to control the camera while the telescope was controlled through the PWI interface. No active guiding was used during the four transit observations. Active guiding is a feature that will be implemented soon and used for future transit photometry observations.

A series of calibration frames were obtained for each of the transit observations and standard photometric reduction procedures were followed to produce our calibrated science images. We then extracted photometry from our science images using the multi-aperture mode of AstroImageJ \citep{2013ascl.soft09001C,2017AJ....153...77C}, which uses simple differential aperture photometry and sky-background subtraction. We then re-centered the apertures on individual stellar centroids in each image using the center-of-light method \citep{2006hca..book.....H}. Details on the aperture size and the comparison stars used to produce the photometry are described in the respective subsections. The choice of comparison stars for each transit observation was based on their counts, trends, and the amount of available stars in the image. The AstroImageJ differential photometry processor automatically removes any comparison star trends by comparing the flux in its aperture to the sum of the flux in all other comparison star apertures \citep{2017AJ....153...77C}.

All fits were performed with EXOFASTv2 \citep[][ Eastman in prep]{2013PASP..125...83E,2017ascl.soft10003E}, which uses a differential evolution Markov chain Monte Carlo to model the stellar system by simultaneously fitting our MINERVA transit, detrending with airmass; the discovery radial velocities from HIRES (HD189733), SOPHIE (WASP-2), and CORALIE (WASP-44 and WASP-45); the spectral energy distribution (SED) using catalog photometry from Tycho \citep{2000AA...355L..27H}, 2MASS \citep{2003yCat.2246....0C}, WISE \citep{2013yCat.2328....0C}, and Gaia \citep{2018AA...616A...1G}; and the MIST stellar evolutionary models \citep{2016ApJS..222....8D,2016ApJ...823..102C}. For each fit, we used used Gaussian priors on T$_{\rm eff}$ and [Fe/H] from the the high resolution spectroscopy in their respective discovery papers, as well as Gaussian priors on the parallax from Gaia DR2, adding 82\,$\mu$ as to correct for the systematic offset found by \citet{2018ApJ...862...61S} and adding the 33\,$\mu$ as uncertainty in their offset in quadrature to the Gaia-reported uncertainty. We applied an upper limit on the V-band extinction from the \citet{2011ApJ...737..103S} dust maps at the location of each target.

The priors and broadband magnitudes we used for each system are summarized in Table~\ref{table:priors}, and the results for each system are summarized in Table~\ref{table:results}. In nearly all cases (unless otherwise noted), our results are consistent with the literature values, and in some cases our uncertainties are smaller, owing primarily to the Gaia constraint on the stellar parameters, but also to the longer baseline between the RVs and our transits.

\startlongtable{
\begin{deluxetable*}{lccccc}
\tablecaption{Priors used for the EXOFASTv2 fitting analysis of the four transiting exoplanets.}
\tablehead{\colhead{~~~Parameter} & \colhead{Description} & \colhead{HD189733} & \colhead{WASP-2} & \colhead{WASP-44} & \colhead{WASP-45}}
\startdata
\vspace{-10pt}
\\\multicolumn{2}{l}{Stellar Parameters:}&\\
                        ~~~~$T_{\rm eff}$\dotfill &                Effective Temperature (K)\dotfill &                                $5050\pm50$ (1)&                                $5200\pm200$ (4)&                                       $5400\pm150$ (5)&                                      $5100\pm200$ (5)\\
                       ~~~~$[{\rm Fe/H}]$\dotfill &                        Metallicity (dex)\dotfill &                        $-0.030\pm0.040$ (1)&                             \dotfill&                                       $0.060\pm0.100$ (5) &                                $0.360\pm0.120$ (5)\\
                                ~~~~$A_V$\dotfill &                  V-band extinction (mag)\dotfill &                         $\leq2.469$ (2)&                         $\leq0.341$ (2)&                         $\leq0.089$ (2)&                         $\leq0.089$ (2)\\
                             ~~~~$\varpi$\dotfill &                           Parallax (mas)\dotfill &                                  $50.651\pm0.048$ (3)&                                   $6.580\pm0.077$ (3)&                         $2.797\pm0.054$ (3)&                              $4.788\pm0.052$ (3)\\
\vspace{-10pt}
\\\multicolumn{2}{l}{Broadband Magnitudes:}&\\
                        ~~~~$B_{T}$\dotfill &                            Tycho $B_{T}$ mag. (6)\dotfill &                                  $8.847^{+0.020}_{-0.016}$&                                   \dotfill&                         \dotfill&                             $13.826\pm0.428$\\
                        ~~~~$V_{T}$\dotfill &                            Tycho $V_{T}$ mag. (6) \dotfill &                                  $7.779^{+0.020}_{-0.010}$&                                   \dotfill&                         \dotfill&                            $11.984\pm0.176$\\
                        ~~~~$J$\dotfill &                                2MASS $J$ mag. (7)\dotfill &                                  $6.073\pm0.030$&                                   $10.166\pm0.030$&                         $11.702\pm0.020$&                            $10.753\pm0.020$\\
                        ~~~~$J$\dotfill &                                2MASS $H$ mag. (7)\dotfill &                                  $5.587\pm0.030$&                                   $9.752\pm0.030$&                         $11.408\pm0.030$&                            $10.365\pm0.030$\\
                        ~~~~$J$\dotfill &                                2MASS $K_{s}$ mag. (7)\dotfill &                                  $5.541\pm0.020$&                                   $9.632\pm0.020$&                         $11.341\pm0.030$&                            $10.294\pm0.020$\\
                        ~~~~$WISE1$\dotfill &                                $WISE1$ mag. (8)\dotfill &                                  $5.289\pm0.154$&                                   $9.582^{+0.030}_{-0.022}$&                         $11.246^{+0.030}_{-0.022}$&                            $10.207^{+0.030}_{-0.022}$\\
                        ~~~~$WISE2$\dotfill &                                $WISE2$ mag. (8)\dotfill &                                  $5.342\pm0.050$&                                   $9.637^{+0.030}_{-0.021}$&                         $11.301^{+0.030}_{-0.021}$&                            $10.275^{+0.030}_{-0.020}$\\
                        ~~~~$WISE3$\dotfill &                                $WISE3$ mag. (8)\dotfill &                                  $5.459^{+0.030}_{-0.013}$&                         $9.546\pm0.038$&                         $11.345\pm0.191$&                            $10.183\pm0.059$\\
                        ~~~~$WISE4$\dotfill &                                $WISE4$ mag. (8)\dotfill &                                  $5.427^{+0.100}_{-0.033}$&                         $8.727\pm0.428$&                            \dotfill&                                   \dotfill\\
                        ~~~~$Gaia$\dotfill &                                $Gaia$ mag. (3)\dotfill &                                  $7.414^{+0.020}_{-0.000}$&                           $11.564^{+0.020}_{-0.001}$&                         $12.896^{+0.020}_{-0.000}$&                            $12.049^{+0.020}_{-0.000}$\\
                        ~~~~$Gaia_{BP}$\dotfill &                                $Gaia_{BP}$ mag. (3)\dotfill &                        $7.913^{+0.020}_{-0.002}$&                         $12.047^{+0.020}_{-0.002}$&                         $13.303^{+0.020}_{-0.002}$&                            $12.515^{+0.020}_{-0.001}$\\
                        ~~~~$Gaia_{RP}$\dotfill &                                $Gaia_{RP}$ mag. (3)\dotfill &                        $6.808^{+0.020}_{-0.002}$&                         $10.946^{+0.020}_{-0.001}$&                         $12.347^{+0.020}_{-0.001}$&                            $11.450^{+0.020}_{-0.001}$\\
\enddata
\tablecomments{References are: (1) \citet{2005AA...444L..15B},
(2) \citet{2011ApJ...737..103S}, (3) \citet{2018AA...616A...1G}, (4) \citet{2008ApJ...677.1324T}, (5) \citet{2017AA...602A.107B}, (6) \citet{2000AA...355L..27H}, (7) \citet{2003yCat.2246....0C}, (8) \citet{2013yCat.2328....0C}.}
\end{deluxetable*}
\label{table:priors}
}

\startlongtable
\begin{deluxetable*}{lccccc}
\tablecaption{Median values and 68\% confidence intervals for the four exoplanetary systems from the MCMC EXOFASTv2 analysis.}
\tablehead{\colhead{~~~Parameter} & \colhead{Description} & \colhead{HD189733} & \colhead{WASP-2} & \colhead{WASP-44} & \colhead{WASP-45}}
\startdata
\multicolumn{2}{l}{Stellar Parameters:}&\smallskip\\
                                ~~~~$M_*$\dotfill &                             Mass (\msun)\dotfill &                         $0.805^{+0.034}_{-0.030}$&                         $0.905^{+0.052}_{-0.049}$&                         $0.929^{+0.053}_{-0.050}$&                         $0.932^{+0.045}_{-0.046}$\\
                                ~~~~$R_*$\dotfill &                           Radius (\rsun)\dotfill &                      $0.7772^{+0.0099}_{-0.0093}$&                         $0.877^{+0.013}_{-0.012}$&                         $0.923^{+0.021}_{-0.020}$&                                   $0.891\pm0.013$\\
                                ~~~~$L_*$\dotfill &                       Luminosity (\lsun)\dotfill &                         $0.355^{+0.014}_{-0.013}$&                         $0.507^{+0.023}_{-0.029}$&                         $0.680^{+0.031}_{-0.029}$&                                   $0.510\pm0.013$\\
                             ~~~~$\rho_*$\dotfill &                            Density (cgs)\dotfill &                            $2.42^{+0.14}_{-0.13}$&                                     $1.89\pm0.13$&                            $1.67^{+0.14}_{-0.13}$&                                     $1.86\pm0.12$\\
                            ~~~~$\log{g}$\dotfill &                    Surface gravity (cgs)\dotfill &                         $4.563^{+0.021}_{-0.020}$&                         $4.509^{+0.026}_{-0.027}$&                                   $4.476\pm0.030$&                         $4.508^{+0.023}_{-0.025}$\\
                        ~~~~$T_{\rm eff}$\dotfill &                Effective Temperature (K)\dotfill &                                $5053^{+46}_{-45}$&                                $5206^{+58}_{-86}$&                                       $5457\pm46$&                                       $5167\pm29$\\
                       ~~~~$[{\rm Fe/H}]$\dotfill &                        Metallicity (dex)\dotfill &                        $-0.003^{+0.031}_{-0.029}$&                            $0.25^{+0.18}_{-0.19}$&                         $0.099^{+0.092}_{-0.089}$&                          $0.388^{+0.090}_{-0.10}$\\
                   ~~~~$[{\rm Fe/H}]_{0}$\dotfill &                     Initial Metallicity \dotfill &                         $0.004^{+0.045}_{-0.044}$&                            $0.24^{+0.15}_{-0.16}$&                         $0.105^{+0.084}_{-0.083}$&                         $0.358^{+0.081}_{-0.091}$\\
                                ~~~~$Age$\dotfill &                                Age (Gyr)\dotfill &                               $6.7^{+4.6}_{-4.2}$&                               $6.2^{+4.7}_{-4.1}$&                               $6.0^{+4.3}_{-3.8}$&                               $5.4^{+4.7}_{-3.5}$\\
                                ~~~~$EEP$\dotfill &                Equal Evolutionary Point \dotfill &                                 $341^{+15}_{-27}$&                                        $348\pm30$&                                 $351^{+36}_{-26}$&                                 $345^{+31}_{-30}$\\
                                ~~~~$A_V$\dotfill &                  V-band extinction (mag)\dotfill &                         $0.127^{+0.059}_{-0.058}$&                         $0.263^{+0.056}_{-0.090}$&                         $0.051^{+0.026}_{-0.033}$&                         $0.022^{+0.013}_{-0.015}$\\
                       ~~~~$\sigma_{SED}$\dotfill &            SED photometry error scaling \dotfill &                            $1.83^{+0.58}_{-0.38}$&                            $1.08^{+0.39}_{-0.25}$&                            $1.51^{+0.59}_{-0.36}$&                            $1.33^{+0.43}_{-0.29}$\\
                             ~~~~$\varpi$\dotfill &                           Parallax (mas)\dotfill &                                  $50.650\pm0.048$&                                   $6.590\pm0.076$&                         $2.805^{+0.052}_{-0.053}$&                         $4.795^{+0.052}_{-0.051}$\\
                                  ~~~~$d$\dotfill &                            Distance (pc)\dotfill &                                  $19.743\pm0.019$&                             $151.7^{+1.8}_{-1.7}$&                             $356.5^{+6.9}_{-6.6}$&                                     $208.5\pm2.2$\\
\\\multicolumn{2}{l}{Planetary Parameters:}&\smallskip\\
                                  ~~~~$P$\dotfill &                            Period (days)\dotfill &             $2.2185788^{+0.0000091}_{-0.0000076}$&                $2.152160^{+0.000025}_{-0.000020}$&                $2.423802^{+0.000032}_{-0.000030}$&                $3.126090^{+0.000037}_{-0.000036}$\\
                                ~~~~$R_P$\dotfill &                             Radius (\rj)\dotfill &                         $1.142^{+0.036}_{-0.034}$&                         $1.117^{+0.025}_{-0.024}$&                         $1.127^{+0.035}_{-0.034}$&                         $0.978^{+0.026}_{-0.024}$\\
                                ~~~~$T_C$\dotfill &            Time of conjunction (\bjdtdb)\dotfill &             $2458334.99057^{+0.00071}_{-0.00073}$&             $2458339.00236^{+0.00042}_{-0.00051}$&                         $2458338.10197\pm0.00036$&             $2458339.14264^{+0.00032}_{-0.00031}$\\
                                ~~~~$T_0$\dotfill &       Optimal $T_C$ (\bjdtdb)\dotfill &             $2458317.24194^{+0.00071}_{-0.00072}$&                         $2458317.48071\pm0.00038$&                         $2458333.25436\pm0.00036$&                         $2458332.89046\pm0.00030$\\
                                  ~~~~$a$\dotfill &                     Semi-major axis (AU)\dotfill &                   $0.03098^{+0.00043}_{-0.00039}$&                   $0.03156^{+0.00060}_{-0.00058}$&                   $0.03446^{+0.00064}_{-0.00063}$&                   $0.04089^{+0.00065}_{-0.00069}$\\
                                  ~~~~$i$\dotfill &                    Inclination (Degrees)\dotfill &                           $85.27^{+0.24}_{-0.23}$&                           $84.38^{+0.27}_{-0.29}$&                           $85.98^{+0.39}_{-0.35}$&                           $84.84^{+0.20}_{-0.24}$\\
                                  ~~~~$e$\dotfill &                            Eccentricity \dotfill &                         $0.024^{+0.026}_{-0.017}$&                          $0.121^{+0.13}_{-0.089}$&                         $0.039^{+0.047}_{-0.028}$&                         $0.048^{+0.034}_{-0.029}$\\
                           ~~~~$\omega_*$\dotfill &         Argument of Periastron (Deg)\dotfill &                                         $27\pm83$&                                $-183^{+19}_{-61}$&                                        $50\pm130$&                                  $58^{+38}_{-43}$\\
                             ~~~~$T_{eq}$\dotfill &              Equilibrium temperature (K)\dotfill &                                       $1220\pm13$&                                $1320^{+20}_{-21}$&                                $1361^{+19}_{-18}$&                                $1163^{+12}_{-11}$\\
                                ~~~~$M_P$\dotfill &                               Mass (\mj)\dotfill &                         $1.130^{+0.047}_{-0.045}$&                         $0.920^{+0.066}_{-0.060}$&                         $0.860^{+0.072}_{-0.068}$&                         $1.018^{+0.046}_{-0.045}$\\
                                  ~~~~$K$\dotfill &                  Radial velocity semi-amplitude (m~s$^{-1}$)\dotfill &                             $202.6^{+6.0}_{-6.2}$&                             $156.4^{+8.1}_{-7.8}$&                             $136.5^{+10.}_{-9.6}$&                                     $147.9\pm4.3$\\
                            ~~~~$\log{K}$\dotfill &                Log of radial velocity semi-amplitude \dotfill &                                   $2.307\pm0.013$&                                   $2.194\pm0.022$&                         $2.135^{+0.031}_{-0.032}$&                         $2.170^{+0.012}_{-0.013}$\\
                            ~~~~$R_P/R_*$\dotfill &       Radius of planet ($R_*$) \dotfill &                                 $0.1510\pm0.0040$&                                 $0.1309\pm0.0020$&                                 $0.1255\pm0.0021$&                      $0.1128^{+0.0022}_{-0.0021}$\\
                              ~~~~$a/R_*$\dotfill &        Semi-major axis ($R_*$) \dotfill &                                     $8.57\pm0.16$&                                     $7.74\pm0.18$&                                     $8.03\pm0.22$&                            $9.87^{+0.20}_{-0.21}$\\
                             ~~~~$\delta$\dotfill &                 Transit depth (fraction)\dotfill &                                 $0.0228\pm0.0012$&                   $0.01714^{+0.00052}_{-0.00053}$&                               $0.01575\pm0.00053$&                   $0.01273^{+0.00051}_{-0.00047}$\\
                               ~~~~$\tau$\dotfill &   Ingress transit duration (days)\dotfill &                      $0.0180^{+0.0016}_{-0.0015}$&                                 $0.0174\pm0.0016$&                      $0.0148^{+0.0015}_{-0.0014}$&                      $0.0236^{+0.0028}_{-0.0024}$\\
                             ~~~~$T_{14}$\dotfill &            Total transit duration (days)\dotfill &                                 $0.0750\pm0.0017$&                                 $0.0751\pm0.0012$&                      $0.0939^{+0.0017}_{-0.0016}$&                                 $0.0695\pm0.0013$\\
                           ~~~~$T_{FWHM}$\dotfill &             FWHM transit duration (days)\dotfill &                      $0.0571^{+0.0022}_{-0.0023}$&                      $0.0577^{+0.0012}_{-0.0013}$&                                 $0.0791\pm0.0011$&                      $0.0459^{+0.0020}_{-0.0024}$\\
                                  ~~~~$b$\dotfill &                Transit Impact parameter \dotfill &                         $0.703^{+0.029}_{-0.033}$&                         $0.737^{+0.023}_{-0.027}$&                         $0.565^{+0.048}_{-0.058}$&                         $0.857^{+0.011}_{-0.013}$\\
                                ~~~~$b_S$\dotfill &                Occultation impact parameter \dotfill &                         $0.710^{+0.031}_{-0.032}$&                         $0.744^{+0.044}_{-0.048}$&                         $0.559^{+0.036}_{-0.039}$&                         $0.914^{+0.059}_{-0.046}$\\
                             ~~~~$\tau_S$\dotfill &   Ingress Occ duration (days)\dotfill &                      $0.0184^{+0.0017}_{-0.0014}$&                      $0.0179^{+0.0029}_{-0.0024}$&                      $0.0147^{+0.0012}_{-0.0011}$&                      $0.0307^{+0.0028}_{-0.0057}$\\
                           ~~~~$T_{S,14}$\dotfill &            Total Occ duration (days)\dotfill &                      $0.0753^{+0.0018}_{-0.0017}$&                      $0.0752^{+0.0012}_{-0.0014}$&                      $0.0939^{+0.0040}_{-0.0036}$&                      $0.0663^{+0.0039}_{-0.0074}$\\
                         ~~~~$T_{S,FWHM}$\dotfill &             FWHM Occ duration (days)\dotfill &                      $0.0570^{+0.0023}_{-0.0025}$&                      $0.0575^{+0.0014}_{-0.0025}$&                      $0.0790^{+0.0036}_{-0.0029}$&                       $0.0331^{+0.011}_{-0.0037}$\\
                ~~~~$\delta_{S,3.6\mu m}$\dotfill &BB Occ depth, 3.6$\mu$m (ppm)\dotfill&                                $1050^{+69}_{-67}$&                                 $983^{+53}_{-50}$&                                 $930^{+58}_{-54}$&                                 $479^{+28}_{-25}$\\
                ~~~~$\delta_{S,4.5\mu m}$\dotfill &BB Occ depth, 4.5$\mu$m (ppm)\dotfill&                                $1575^{+99}_{-95}$&                                $1413^{+67}_{-64}$&                                $1320^{+74}_{-70}$&                                 $743^{+40}_{-36}$\\
                             ~~~~$\rho_P$\dotfill &                            Density (cgs)\dotfill &                          $0.939^{+0.10}_{-0.090}$&                         $0.820^{+0.085}_{-0.076}$&                         $0.746^{+0.096}_{-0.087}$&                                     $1.35\pm0.12$\\
                             ~~~~$logg_P$\dotfill &                         Surface gravity \dotfill &                         $3.331^{+0.033}_{-0.032}$&                         $3.262^{+0.037}_{-0.036}$&                         $3.225^{+0.044}_{-0.045}$&                         $3.421^{+0.029}_{-0.030}$\\
                             ~~~~$\Theta$\dotfill &                         Safronov Number \dotfill &                      $0.0759^{+0.0034}_{-0.0033}$&                      $0.0573^{+0.0037}_{-0.0030}$&                      $0.0566^{+0.0045}_{-0.0043}$&                                 $0.0912\pm0.0036$\\
                              ~~~~$\fave$\dotfill &                 Inc Flux (\fluxcgs)\dotfill &                         $0.502^{+0.022}_{-0.021}$&                         $0.672^{+0.046}_{-0.048}$&                         $0.776^{+0.044}_{-0.040}$&                         $0.414^{+0.016}_{-0.015}$\\
                                ~~~~$T_P$\dotfill &             Time of Periastron (\bjdtdb)\dotfill &                      $2458334.64^{+0.51}_{-0.50}$&                      $2458337.23^{+0.12}_{-0.32}$&                      $2458337.93^{+0.83}_{-0.82}$&                      $2458338.90^{+0.30}_{-0.36}$\\
                                ~~~~$T_S$\dotfill &                Time of Occultation (\bjdtdb)\dotfill &                   $2458336.112^{+0.039}_{-0.021}$&                      $2458337.78^{+0.15}_{-0.19}$&                   $2458339.314^{+0.057}_{-0.054}$&                   $2458340.744^{+0.056}_{-0.044}$\\
                                ~~~~$T_A$\dotfill &         Time of Asc Node (\bjdtdb)\dotfill &                   $2458334.445^{+0.029}_{-0.016}$&                    $2458338.409^{+0.070}_{-0.13}$&                   $2458337.496^{+0.044}_{-0.042}$&                   $2458341.541^{+0.045}_{-0.044}$\\
                                ~~~~$T_D$\dotfill &        Time of Desc Node (\bjdtdb)\dotfill &                   $2458335.548^{+0.020}_{-0.015}$&                   $2458339.456^{+0.082}_{-0.076}$&                             $2458338.708\pm0.036$&                   $2458339.913^{+0.031}_{-0.042}$\\
                    ~~~~$e\cos{\omega_*}$\dotfill &                                         \dotfill &                         $0.009^{+0.027}_{-0.015}$&                           $-0.10^{+0.11}_{-0.14}$&                         $0.000^{+0.037}_{-0.035}$&                         $0.019^{+0.028}_{-0.022}$\\
                    ~~~~$e\sin{\omega_*}$\dotfill &                                         \dotfill &                         $0.003^{+0.023}_{-0.016}$&                                   $0.004\pm0.040$&                                  $-0.000\pm0.037$&                         $0.032^{+0.036}_{-0.030}$\\
                          ~~~~$M_P\sin i$\dotfill &                       Minimum mass (\mj)\dotfill &                         $1.126^{+0.047}_{-0.045}$&                         $0.916^{+0.066}_{-0.059}$&                         $0.858^{+0.072}_{-0.068}$&                         $1.014^{+0.046}_{-0.045}$\\
                            ~~~~$M_P/M_*$\dotfill &                              Mass ratio \dotfill &                             $0.001338\pm0.000044$&                $0.000968^{+0.000061}_{-0.000049}$&                $0.000884^{+0.000067}_{-0.000064}$&                $0.001043^{+0.000035}_{-0.000034}$\\
                              ~~~~$d/R_*$\dotfill &               Separation at mid transit \dotfill &                            $8.52^{+0.23}_{-0.25}$&                            $7.52^{+0.36}_{-0.39}$&                            $8.03^{+0.41}_{-0.42}$&                            $9.54^{+0.41}_{-0.48}$\\
                                ~~~~$P_T$\dotfill &       A priori non-grazing tran prob \dotfill &                      $0.0996^{+0.0031}_{-0.0027}$&                      $0.1156^{+0.0065}_{-0.0053}$&                      $0.1089^{+0.0060}_{-0.0053}$&                      $0.0930^{+0.0050}_{-0.0039}$\\
                            ~~~~$P_{T,G}$\dotfill &                   A priori transit prob \dotfill &                      $0.1351^{+0.0041}_{-0.0037}$&                      $0.1505^{+0.0082}_{-0.0069}$&                      $0.1402^{+0.0077}_{-0.0067}$&                      $0.1167^{+0.0061}_{-0.0048}$\\
                                ~~~~$P_S$\dotfill &       A priori non-grazing occ prob \dotfill &                      $0.0985^{+0.0025}_{-0.0026}$&                      $0.1134^{+0.0092}_{-0.0045}$&                      $0.1091^{+0.0046}_{-0.0043}$&                      $0.0871^{+0.0023}_{-0.0025}$\\
                            ~~~~$P_{S,G}$\dotfill &                   A priori occ prob \dotfill &                                 $0.1336\pm0.0037$&                       $0.1476^{+0.012}_{-0.0064}$&                      $0.1404^{+0.0063}_{-0.0059}$&                      $0.1093^{+0.0032}_{-0.0034}$\\
\\\multicolumn{2}{l}{Wavelength Parameters:}&z'&r'&r'&r'\smallskip\\
                              ~~~~$u_{1}$\dotfill &             linear limb-darkening coeff \dotfill &                                   $0.347\pm0.050$&                         $0.525^{+0.053}_{-0.054}$&                                   $0.448\pm0.048$&                                   $0.545\pm0.051$\\
                              ~~~~$u_{2}$\dotfill &          quadratic limb-darkening coeff \dotfill &                                   $0.220\pm0.050$&                                   $0.188\pm0.052$&                                   $0.230\pm0.050$&                         $0.180^{+0.049}_{-0.050}$\\
\smallskip\\\multicolumn{2}{l}{Telescope Parameters:}&HIRES&SOPHIE&CORALIE&CORALIE\smallskip\\
                   ~~~~$\gamma_{\rm rel}$\dotfill &                 Relative Radial Velocity Offset (m~s$^{-1}$)\dotfill &                             $-14.8^{+4.3}_{-4.4}$&                              $-27857^{+18}_{-17}$&                           $-4045.1^{+7.0}_{-6.2}$&                                    $4548.1\pm3.7$\\
                           ~~~~$\sigma_J$\dotfill &                          Radial Velocity Jitter (m~s$^{-1}$)\dotfill &                              $16.5^{+4.6}_{-3.1}$&                                  $11^{+16}_{-12}$&                              $0.00^{+19}_{-0.00}$&                             $0.00^{+10.}_{-0.00}$\\
                         ~~~~$\sigma_J^2$\dotfill &                      Radial Velocity Jitter Variance \dotfill &                                $270^{+170}_{-94}$&                               $140^{+600}_{-180}$&                               $-20^{+400}_{-200}$&                                 $-7^{+110}_{-50}$\\
\\\multicolumn{2}{l}{Transit Parameters:}& UT 2018-08-04 (z') & UT 2018-08-08 (r') & UT 2018-08-07 (r') & UT 2018-08-08 (r')\smallskip\\
                         ~~~~$\sigma^{2}$\dotfill &                          Added Variance \dotfill &             $0.0001122^{+0.0000075}_{-0.0000067}$&          $0.00000110^{+0.00000030}_{-0.00000027}$&          $0.00000006^{+0.00000033}_{-0.00000025}$&         $-0.00000119^{+0.00000015}_{-0.00000013}$\\
                                ~~~~$F_0$\dotfill &                           Baseline flux \dotfill &                               $0.99951\pm0.00073$&                               $0.99997\pm0.00017$&                   $1.00005^{+0.00026}_{-0.00025}$&                               $1.00001\pm0.00011$\\
\enddata
\label{table:results}
\end{deluxetable*}

\subsection{Transit Observation of WASP-2b}
We observed the transit of WASP-2b \citep{2007MNRAS.375..951C}, starting photometric observations on the evening of 2018 August 8 at approximately 10:30~UT, and continuing observing until 13:15~UT, obtaining 304 science frames with a cadence of 35\,s. WASP-2 is a moderately faint ($V=11.98$) K1V spectral type star hosting a hot Jupiter with an orbital period of $P=2.15$\,days \citep{2007MNRAS.375..951C}. The sky was photometric with clear conditions during the transit observations and seeing of 3.5\arcsec.

From our calibrated science frames, we extracted differential photometry using the AstroImageJ reduction pipeline by first selecting an aperture of 15 pixels (9.15\arcsec) and a sky annulus with an inner radius of 30 pixels and an outer radius of 40 pixels around four stars, including WASP-2b.

\begin{figure}
	\centering
	\includegraphics[width=0.6\linewidth]{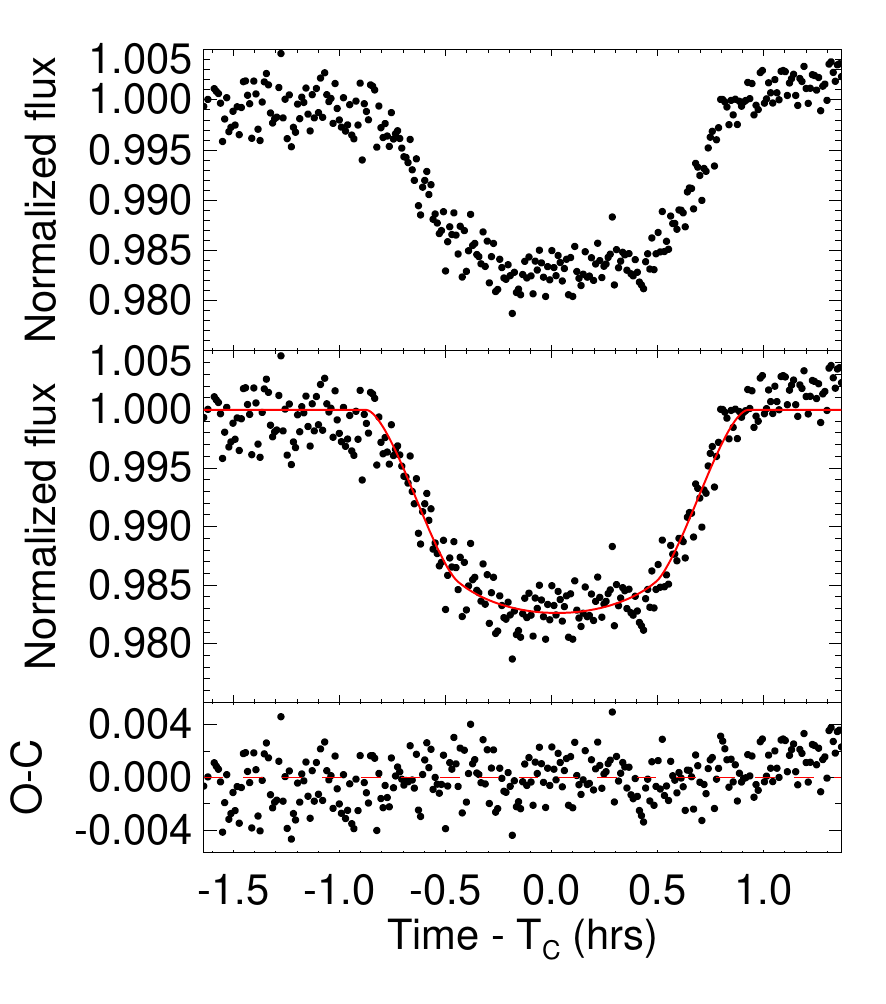}
	\caption{Transit light curve data for WASP-2b; the top light curve is normalized but not de-trended and the bottom one is normalized and de-trended with a best-fit model overplotted. The bottom panel shows the residuals to the best-fit model. $T_{C}$ is the mid-transit time.}
	\label{fig:wasp2b}
\end{figure}

Once we extracted the photometry, we fit the data using EXOFASTv2 to produce the final light curve of WASP-2b. Figure ~\ref{fig:wasp2b} shows the raw light curve, the best-fit transit model with the de-trended light curve, and the residuals from the fit. The RMS scatter of the residuals from the fit to our light curve is 0.95\,mmag or 950 parts per million (ppm).

Our results for WASP-2b are in general agreement with those of \citet{2007MNRAS.375..951C}. The mid-transit time we measured ($T_C=2458339.00342_{-0.00051}^{+0.00042}$) is -0.00876\,days (12.6\,min) earlier than the predicted time of $T_C=2458339.01112\pm0.0092$. This is in agreement with the \citet{2007MNRAS.375..951C} predicted mid-transit time when the accumulated uncertainty from 2020 orbital cycles since the last published ephemeris has been taken into account. Our measurement updates the published ephemeris, and extends the time baseline of WASP-2b transit photometry to approximately 12 years. 

The orbital period we measured is $P=2.152160_{-0.000020}^{+0.000025}$ days, which does not agree to within $3\sigma$ of the published period of $P=2.152226\pm0.000004$\,days (difference of 0.000066\,days or $\sim3.3\sigma$). The source of this discrepancy is not clear and \citet{2017MNRAS.472.3871T} found no evidence for transit timing variations. The results for our other parameters are in agreement with those of \citet{2007MNRAS.375..951C}. 

\subsection{Transit Observation of WASP-44b}
Transit observations of WASP-44b were started on the night of 2018 August 7 at approximately 12:35~UT, nearly an hour before transit ingress. We continued to observe WASP-44b for about an hour after egress, observing until 16:10~UT, collecting 62 science frames with a cadence of 205\,s. WASP-44b is a relatively faint ($V=12.9$) G8V spectral type star that hosts a hot Jupiter with an orbital period of $P=2.42$\,days \citep{2012MNRAS.422.1988A}. Our observations were done under clear skies and seeing of 4.1\arcsec.

\begin{figure}
	\centering
	\includegraphics[width=0.6\linewidth]{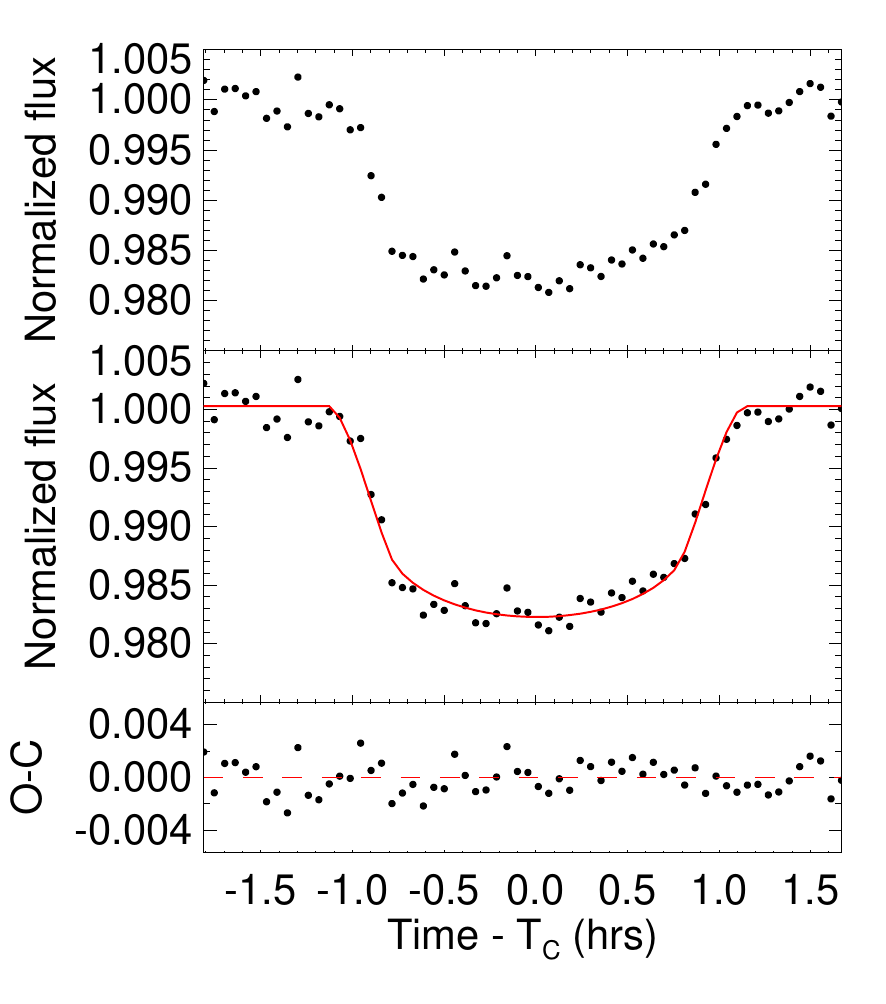}
	\caption{Same as Figure~\ref{fig:wasp2b}, but for the light curve of WASP-44b.}
	\label{fig:wasp44b}
\end{figure}

Photometry was extracted using the AstroImageJ reduction pipeline following the same procedure as detailed for WASP-2b and using the same size aperture and sky annulus around our target and comparison stars. We then fit the data using EXOFASTv2 to produce the final light curve of WASP-44b (Figure~\ref{fig:wasp44b}). The RMS scatter of the residuals from the fit to our light curve is 1.18\,mmag (1180 ppm). The best-fit transit model with the de-trended light curve and the residuals from the fit are shown in Figure~\ref{fig:wasp44b}. Table~\ref{table:results} lists the resulting median parameter values and $1\sigma$ uncertainties.

Overall our results for WASP-44b are in good agreement with those of \citet{2012MNRAS.422.1988A} to within $1\sigma$. The mid-transit time we measured ($T_C=2458338.10197\pm0.00036$) is 0.0089\,days or 12.8\,min later than the predicted time of of $T_C=2458338.09307\pm0.01040$, but is in agreement with \citet{2012MNRAS.422.1988A} when the accumulated uncertainties from 1198 orbital cycles since the last ephemeris have been taken into account for the mid-transit time uncertainty. Thus our measurements do agree with the predicted time and suggest that there is no significant deviation in the transit time. We have now provided an update to the published ephemeris that extends the time baseline of WASP-44b transit photometry to approximately eight years.

\subsection{Transit Observation of WASP-45b}
On the night of 2018 August 8, we observed another transiting planet, WASP-45b, after observations of WASP-2b had finished. For this transit, observing commenced at 13:55~UT (just over an hour before ingress) and continued until 17:20~UT (around two hours after egress). In total we obtained 219 science frames with a cadence of 55\,s. WASP-45 is a moderately faint ($V=12.0$) K2V spectral type star hosting a hot Jupiter with an orbital period of $P=3.13$\,days \citep{2012MNRAS.422.1988A}. WASP-45b was observed under similarly clear skies; however, the seeing was worse (6.1\arcsec).

\begin{figure}
	\centering
	\includegraphics[width=0.6\linewidth]{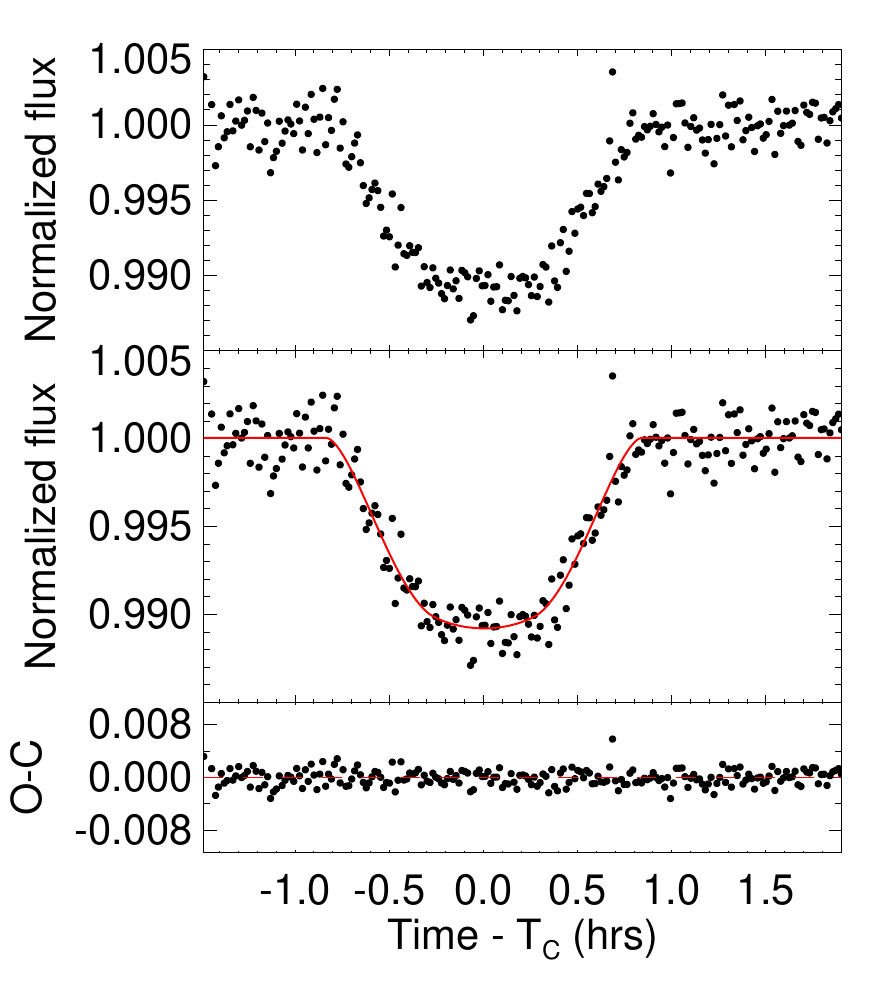}
	\caption{Same as Figure~\ref{fig:wasp2b}, but for the light curve of WASP-45b.}
	\label{fig:wasp45b}
\end{figure}

We followed the same overall procedure for extracting photometry as described in Section~\ref{PrecisionPhotometry}. In total nine comparison stars were used for deriving the photometry. The raw light curve along with the best-fit model with the de-trended light curve and the residuals from the fit are shown in Figure~\ref{fig:wasp45b}. The RMS scatter from our best-fit light curve model is 1.2\,mmag (1200 ppm).

Our results for WASP-45b are generally in good agreement with those of \citet{2012MNRAS.422.1988A}. The mid-transit time we measured ($T_C=2458339.14264_{-0.00031}^{+0.00032}$) is -0.00982\,days (-14.1\,min) earlier than the predicted time of $T_C=2458339.15246\pm0.00058$ from \citet{2012MNRAS.422.1988A} but is in agreement with their result when the uncertainty on the orbital period is taken into account after 927 orbital periods. Our transit observation now extends the published ephemeris to a time baseline of approximately eight years.

There are 1 to $1.5\sigma$ discrepancies seen in $T_{14}$, $i$, $R_{*}$, and $R_{P}$. The origin of this discrepancy is not understood but could potentially be from stellar activity as \citet{2012MNRAS.422.1988A} reported WASP-45 to be chromospherically active.

\subsection{Transit Observation of HD 189733b}
We observed the transit of HD\,189733b on the night of UT 2018 August 4 with photometric observations starting at approximately 10:00UT (about one hour before ingress). We continued observing the target until approximately 30 minutes after egress (13:15UT), obtaining 517 observations with a cadence of $\sim7$\,s. HD\,189733 is a very bright ($V=7.6$) K2 spectral-type star hosting a hot Jupiter with an orbital period of $P=2.22$ \citep{2005A&A...444L..15B}. The observations were conducted under clear skies and seeing of 2.6\arcsec.

\begin{figure}
	\centering
	\includegraphics[width=0.6\linewidth]{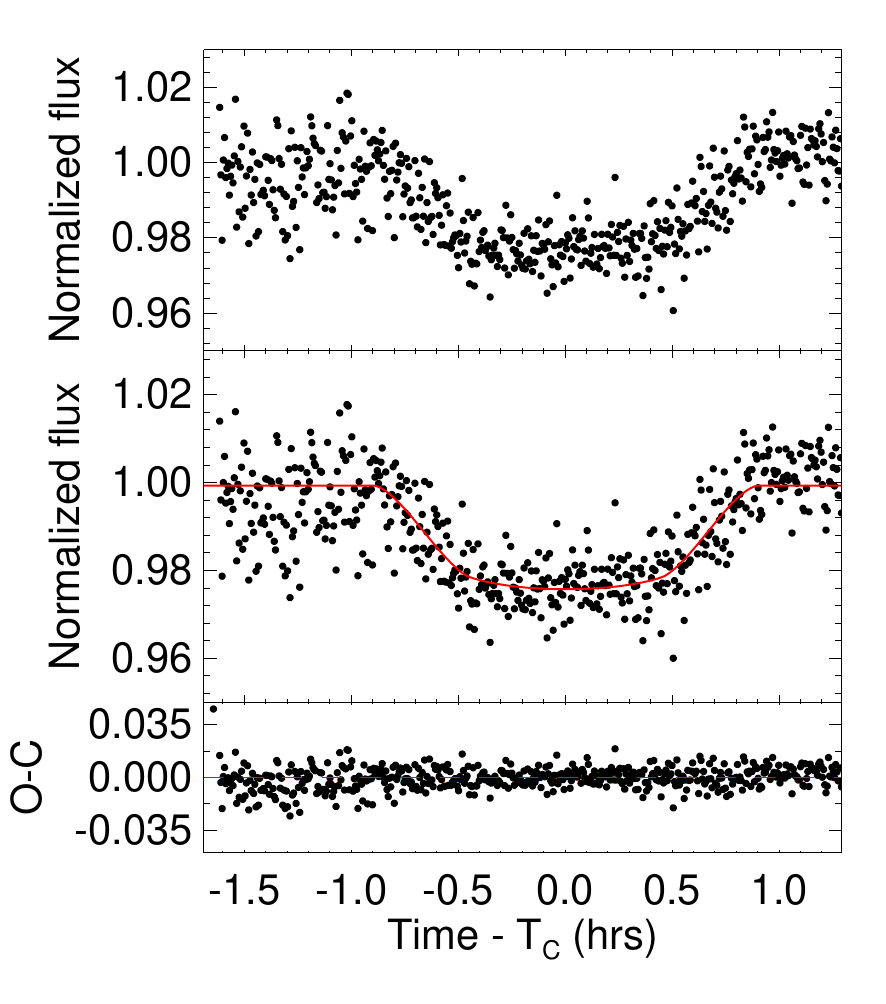}
	\caption{Same as Figure~\ref{fig:wasp2b}, but for the light curve of HD 189733b. 
	}
	\label{fig:HD189733b}
\end{figure}

Following the procedure described in Section~\ref{PrecisionPhotometry}, we extracted photometry using the AstroImageJ by first selecting an aperture of 20 pixels (12.3\arcsec) and a sky annulus with an inner radius of 35 pixels and an outer radius of 45 pixels around 14 stars, including HD\,189733. The raw and de-trended light curve with the best-fit model along with the residuals from the fit are shown in Figure~\ref{fig:HD189733b}. The RMS scatter of the residuals from the fit to our light curve is 5.2\,mmag (5200 ppm).

Our results for HD\,189733b are in good agreement with those of \citet{2005A&A...444L..15B}, \citet{2006ApJ...650.1160B}, \citet{2010MNRAS.408.1689S}, and \citet{2015MNRAS.450.3101B}. The mid-transit time we measured ($T_C=2458334.99057_{-0.00073}^{+0.00071}$) is -0.0024\,days (-3.5\,min) earlier than the predicted time of $T_C=2458334.9930000\pm0.0000088$ from \citet{2015MNRAS.450.3101B} but is in agreement with their result when the uncertainty on the orbital period is taken into account after 1974 orbital periods. HD\,189733b now has an ephemeris baseline of approximately 13 years.

\section{Conclusions and Future Work}
\label{Conclusions}

We have built a dedicated observatory for high-precision photometric and spectroscopic observations of exoplanetary systems, primarily in support of the NASA {\textit {TESS}} mission.  {\sc Minerva}-Australis is the only Southern Hemisphere facility with such capabilities that is fully dedicated to {\textit {TESS}} follow-up. 

In this work, we have presented initial photometric science demonstration results. Based on these results, we expect to contribute high-precision ($\leq1.0$\,mmag) photometry for \textit{TESS} targets brighter than $V=13$ with exposures of under five minutes. This is comparable to the {\sc Minerva} North photometric commissioning results \citep{2015JATIS...1b7002S} and to other ground-based follow-up programs such as HATSouth carried out on the Las Cumbres Observatory 1\,m telescopes, the Danish 1.54\,m telescope, the Chilean-Hungarian Automated 0.7\,m Telescope, and Perth Exoplanet Survey 0.3\,m Telescope which obtained photometric precision of 0.9--3.0\,mmag for $V\leq13$ targets \citep[see, e.g.,][]{2019AJ....157...55H}. At the time of writing, the KiwiSpec spectrograph is being commissioned.  In a forthcoming paper (D. J. Wright et al. in preparation), we will fully describe the acquisition and analysis of spectroscopic data at {\sc Minerva}-Australis, and we will present data demonstrating our radial velocity precision on standard stars and known exoplanets. 

We are also in the process of enabling fully autonomous operations of the telescope array by customizing the existing automation software, {\sc Minerva} Robotic Software \citep{2015JATIS...1b7002S}, developed for the {\sc Minerva} North array to suit our specific scientific goals and operational requirements. 

In future, we anticipate adding very high-cadence photometric capability, as each telescope can rapidly switch between photometric and spectroscopic modes via use of the two Nasmyth ports.  With photometric cadence of up to 20 Hz, {\sc Minerva}-Australis will be able to capture occultation events of small solar system bodies \citep[e.g.][]{occ1,occ2,occ3}, allowing us to pursue time-critical target-of-opportunity research, in addition to pursuing our core goals in exoplanetary science. 

\section{Acknowledgements}

This research was supported by the Australian Government through the Australian Research Council's Discovery Projects funding scheme (project DP180100972).  {\sc Minerva}-Australis hardware is funded in part by the Australian government through the Australian Research Council, LIEF grants LE160100001.  We acknowledge support from the Mount Cuba Astronomical Foundation. H.Z. is also grateful to the support from the Natural Science Foundation of China (NSFC grants 11673011, 11333002). P.P. acknowledges support from the the National Science Foundation (Astronomy and Astrophysics grant 1716202) and George Mason University start-up funds.

The {\sc Minerva} in the Northern hemisphere, which laid the groundwork for our installation, is made possible by generous contributions from its collaborating institutions and Mt. Cuba Astronomical Foundation, The David \& Lucile Packard Foundation, National Aeronautics and Space Administration (EPSCOR grant NNX13AM97A), The Australian Research Council (LIEF grant LE140100050), and the National Science Foundation (grants 1516242 and 1608203).

\pagebreak

\bibliography{references}

\end{document}